\documentclass[a4paper,fleqn, sort&compress]{cas-dc}

 \usepackage[numbers]{natbib}
\usepackage{graphicx}
\usepackage{url}
\usepackage{booktabs}
\usepackage{multirow}
\usepackage{subfigure}
\usepackage{amsmath}
\usepackage{amssymb}
\usepackage{color}
\usepackage{physics}
\usepackage{enumitem}
\usepackage{tikz}
\newcommand*\mycirc[1]{%
\begin{tikzpicture}[baseline=(C.base)]
\node[draw,circle,inner sep=1pt,minimum size=2ex](C) {#1};
\end{tikzpicture}}
\newcommand{\di}{\text{d}}

\begin{document}
\let\WriteBookmarks\relax
\def\floatpagepagefraction{1}
\def\textpagefraction{.001}

\shorttitle{Time-sharing Orbit Jump and Energy Harvesting in Nonlinear Piezoelectric Energy Harvesters}

\shortauthors{B. Zhao et~al.}

\title [mode = title]{Time-sharing Orbit Jump and Energy Harvesting in Nonlinear Piezoelectric Energy Harvesters Using a Synchronous Switch Circuit}                      



%
\author[1,4]{Bao Zhao}[orcid=0000-0002-9689-7742]

\credit{Conceptualization, Formal analysis, Investigation, Methodology, Resources, Writing – original draft}
\affiliation[1]{organization={School of Information Science and Technology, ShanghaiTech University},
    city={Shanghai},
    postcode={201210}, 
    country={China}}

\author[2]{Jiahua Wang}
\affiliation[2]{organization={Department of Mechanical and Automation Engineering, The Chinese University of Hong Kong},
    addressline={Shatin, NT}, 
    city={Hong Kong},
    country={China}}
\credit{Conceptualization, Validation, Investigation, Writing – original draft}

\author[3]{Guobiao Hu}[orcid=0000-0002-1288-7564]
\affiliation[3]{organization={Thrust of the Internet of Things, The Hong Kong University of Science and Technology (Guangzhou)},
    addressline={Nansha}, 
    city={Guangzhou},
    postcode={511400}, 
    country={China}}
\credit{Investigation, Writing – review \& editing}

\author[4]{Andrea Colombi}[orcid=0000-0003-2480-978X]
\affiliation[4]{organization={Department of Civil, Environmental, and Geomatic Engineering, ETH Zürich},
    city={Zürich},
    postcode={8093}, 
    country={Switzerland}}
\credit{Writing – review \& editing}

\author[2]{Wei-Hsin Liao}[orcid=0000-0001-7221-5906]
\ead{whliao@cuhk.edu.hk}
\credit{Supervision, Writing – review \& editing}
\cormark[1]

\author[1]{Junrui Liang}[orcid=0000-0003-2685-5587]
\ead{liangjr@shanghaitech.edu.cn}
\credit{Conceptualization, Supervision, Resources, Writing – review \& editing}
\cormark[1]

\cortext[cor1]{Corresponding author}


\nonumnote{This paper is an extending version of \cite{wang2019orbit}, which was presented at the 2019 ASME International Design Engineering Technical Conferences and Computers and Information in Engineering Conference. (Bao Zhao and Jiahua Wang contributed equally to this work.)}


\begin{abstract}
    Nonlinearity has enabled energy harvesting to advance towards higher power output and broader bandwidth in monostable, bistable, and multistable systems. However, operating in the preferable high-energy orbit (HEO) rather than the low-energy orbit (LEO) for making such advancement has restricted their applications. Based on a monostable nonlinear system, this paper proposes a self-contained solution for time-sharing orbit jump and energy harvesting. The joint dynamics of an electromechanical assembly consisting of a nonlinear energy harvester and a switched-mode piezoelectric interface circuit for high-capability energy harvesting is studied. The proposed solution is carried out by utilizing a cutting-edge switched-mode bidirectional energy conversion circuit (BECC), which enables time-sharing dual functions of energy harvesting and vibration exciting.  A theoretical model is established based on impedance analysis and multiple time scales method to analyze the stability, frequency response, and phase evolution of the autonomous and nonautonomous nonlinear energy harvesting systems. In particular, the detailed dynamics for the orbit jumps with the vibration exciting mode of BECC are studied. Experiments are performed to validate the full-hysteresis-range orbit jumps with the monostable nonlinear energy harvester. The harvested power after orbit jumps yields a nine-fold increase, compensating for the energy consumption under vibration exciting mode quickly. The proposed solution also refrains the system from extra mechanical or electrical energy sources for orbit jumps, which leads to the first self-contained solution for simultaneous energy harvesting and orbit jump in nonlinear piezoelectric energy harvesting. This work enhances the practical utility of nonlinear energy harvesting technologies toward engineering applications. 
\end{abstract}



\maketitle

\section{Introduction}

Energy harvesting has been widely investigated over the last two decades as a potential solution for powering wireless sensor nodes in Internet of Things (IoT) applications. It enables the system to collect and convert energy, which opens up opportunities for self-sustaining systems \cite{shaikh2016energy}. According to application scenarios, different technologies may target solar energy, thermal energy, or vibration energy \cite{prauzek2018energy}. Among them, vibration energy harvesting draws massive attention due to its easy accessibility in the ambient environment. The most investigated electromechanical transduction mechanisms include the piezoelectric, electromagnetic, and electrostatic ones \cite{wei2017comprehensive}. 

This paper focuses on piezoelectric energy harvesting (PEH) as a result of its high power density and compatibility in small-scale systems. Previous efforts have been made to improve power output and broaden the bandwidth of energy harvesters \cite{zhao2022circuit}. 

For one thing, mechanically, multi-resonance \cite{zhao2022graded} and frequency-self-tuning \cite{wang2017frequency} structures were developed to create a resonance condition for broadening the effective energy harvesting bandwidth. However, their power density and overall capability are inferior, compared with the nonlinear mechanical design \cite{fang2022multistability}. To form a nonlinear energy harvester, attractive or repulsive forces, asymmetric geometry, and post-buckling configuration are usually utilized \cite{zou2019mechanical}. The caused nonlinearity expands the working bandwidth and may increase the power output, owing to the hardening or softening effects.
Nevertheless, at the same time, the consequent hysteresis creates multiple root branches where the high-energy orbit (HEO) and the low-energy orbit (LEO) coincide \cite{guckenheimer2013nonlinear}. For the objective of energy harvesting, the oscillation on HEO with a much higher power output is preferable. The studies of orbit jump in the energy harvesting field started by Erturk et al. \cite{Erturk2011}, Sebald et al. \cite{Sebald2011}, and Masuda et al. \cite{masuda2011vibration} in the early 2010s by methods of external impacts by hands, high voltage excitation with piezoelectric actuators, and negative resistance, respectively. Since then, different methods have been proposed for this target.
The majority of the orbit jump methods utilize the system parameter tuning or perturbation strategies to force or perturb the oscillator out of its original LEO and seek the path to the HEO in the phase space, including negative impedance \cite{lan2017obtaining,Kitamura2018}, load perturbation \cite{wang2019attaining}, stiffness modulation \cite{yan2019low}, buckle level modulation \cite{huguet2019orbit,huang2020high}. 
Meanwhile, other methods enable extra energy injected into the oscillator, which also facilitates the orbit jumps, including energy transfer by projectile impact \cite{Zhou2015}, piezoelectric voltage excitation \cite{Mallick2016}, magnetic plucking with bistable energy harvesters \cite{Fu2017}, attractor selection \cite{Udani2017}, sliding mode control methods \cite{haji2018robust,zhang2021high}, and excitation tuning \cite{yu2020capture}. These methods were analytically and experimentally demonstrated, but these solutions are still tough to be implemented outside the laboratory since most of them rely on extra devices or energy sources to adjust their states. In addition, the energy or control cost of tuning the parameter may outweigh the overall outcome. In general, there are two common issues for carrying out these existing methods.
\begin{itemize}
    \item An extra energy source, which is capable of intensive mechanical drive output, is required.
    \item An intensive energy injection, if introduced at an improper phase, may induce much negative work and reduces the actuating efficiency.
\end{itemize}
Therefore, more controllable, self-contained, and efficient energy injection methods for orbit jumps are still in demand to practically enhance the nonlinear energy harvester performance and implement the energy harvester in the field. 

For another, electrically, many piezoelectric interface circuits are proposed. The linear AC energy harvesting circuit of a pure resistive load has reactive power under some phase ranges \cite{Liang2012}. To better improve the power factor, more advanced nonlinear AC-DC circuits, required by most electronic devices, were developed. The standard energy harvesting circuit utilizes the rectifier and smoothing capacitor to output a DC voltage to the load \cite{Ottman2002}. Afterward, different charge manipulation approaches were employed to increase the power output further. The parallel or series synchronized switch harvesting on inductor (SSHI) circuits flips the piezoelectric voltage at displacement extreme over an inductor. They enlarge the harvested power by several folds \cite{Guyomar2005, Lien2010}. The synchronized electric charge extraction (SECE) circuit enhances the power output and has a load-independent feature \cite{Chen2019}. Combining the SSHI and SECE concepts created the double synchronized switch harvesting circuit, which allows control of the extraction process and increase of voltage \cite{Lallart2008}. The synchronized triple bias-flip (S3BF) circuit \cite{Liang2018} is proposed to reduce the switching loss and increase the voltage level by introducing three successive voltage bias-flips during each voltage-flip process. Based on this circuit, Zhao et al. \cite{zhao2020series, zhao2021bidirectional} proposed the bidirectional energy conversion circuit (BECC) by removing the bridge rectifier. The BECC enables the dual functions of energy harvesting and vibration exciting by using the same circuit in a time-sharing manner.
Compared with the applications in linear energy harvesting systems, some recent works have investigated these interface circuits with nonlinear energy harvesters to explore the joint dynamics \cite{Dai2018, Lan2018, lai2023energy}. Their results showed that the circuit might further extend the systems’ bandwidth and increase the power output \cite{zhao2022circuit}. But how to ensure a vibration in the HEO in the inherent hysteresis range has not been addressed. As for the S3BF circuit, its interaction with a nonlinear energy harvester is not well-studied yet. 

Although many efforts have been taken to enhance the power output and broaden the bandwidth of energy harvesters, general modeling methods covering both mechanical dynamics and electrically induced dynamics are quite deficient. Besides, present orbit jumps heavily rely on extra intensive mechanical or electrical energy sources, which is not realistic in practical applications. This work uses an impedance model and multiple time scales method to investigate the joint dynamics of a monostable energy harvester connected with the BECC circuit. Based on the dual functions of this circuit, this research proposes an efficient vibration exciting strategy for orbit jumps to enhance the harvested power and bandwidth for nonlinear energy harvester.

This paper is organized as follows. Following the introduction, the integration of the monostable energy harvester and BECC is presented in Section \ref{sec:over}. The equivalent impedance model and electrically induced parameters are defined. In Section \ref{sec:dynamics}, the dynamics of the integrated system are investigated. Firstly, the stability analysis of an autonomous case is studied to reveal the effects of different operation modes on the dynamics of the nonlinear harvester. Next, the nonautonomous systems' frequency response and state-space phase evolution are performed with a multiple time scales method. The detailed steps for orbit jump from low-energy orbits to high-energy orbits are analyzed. Section \ref{sec:exp} presents the experimental results for orbit jumps of the nonlinear energy harvesting system with BECC. An energy evaluation is conducted to quantify the overall energy consumption during orbit jumps. Section \ref{sec:dis} discusses the extension of this orbit jump solution further to a bistable energy harvesting system. Finally, the conclusions are drawn in Section \ref{sec:conclusion}.

\section{System Overview}
\label{sec:over}
\subsection{Nonlinear PEH System using BECC}
\begin{figure}[!t]
    \centering
    \includegraphics[width=\columnwidth,page=1]{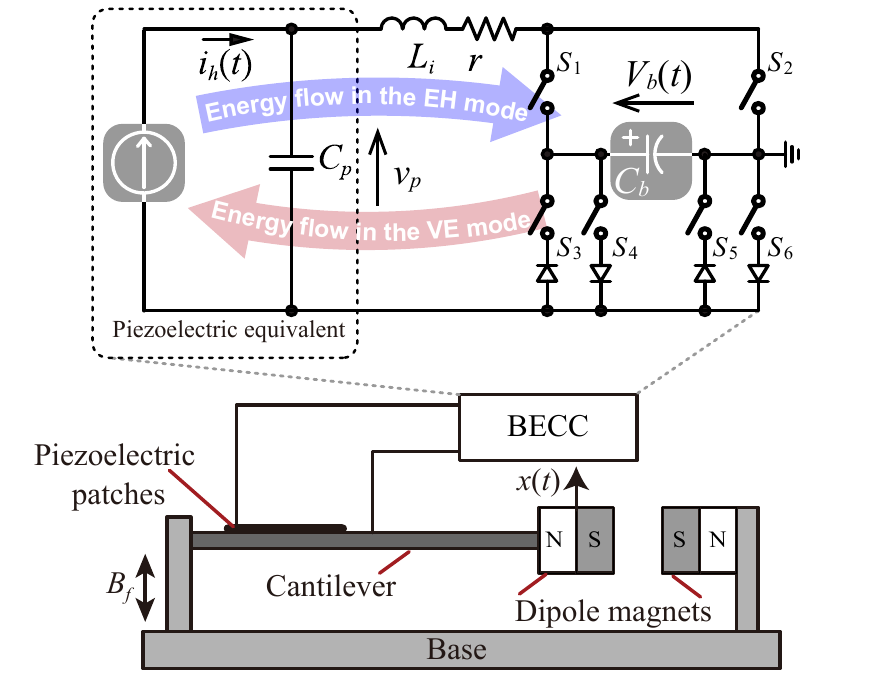}
    \caption{System overview of a nonlinear piezoelectric energy harvester with the bidirectional energy conversion circuit (BECC). }
   \label{fig:system}
 \end{figure}

As Fig. \ref{fig:system} shows, a nonlinear piezoelectric energy harvester is considered. The nonlinearity is introduced by the repelling force of two opposite magnets. A monostable oscillator is achieved by tuning the distance between magnets. The BECC interface circuit, to be explained in the Subsection \ref{subsec:circuit}, is connected with the piezoelectric energy harvester. The equations of motion for the integrated system can be represented as \cite{Liang2012}:
\begin{equation}
    \left\{
    \begin{aligned}
        &M \ddot{x}+C \dot{x}+K x-K_{1} x+K_{2} x^{3}+\alpha_e v_p=B_{f} \cos \left(\omega t\right) \\
    	&\alpha_e \dot{x}-C_p \dot{v}_p-i_p=0
    \end{aligned}
    \right.
	\label{eq:eom}
\end{equation}
where $M$, $C$, and $K$ are equivalent mass, damping, and stiffness of the cantilever beam, while $K_1$ and $K_2$ denote the coefficients of the nonlinear stiffness caused by magnets, respectively. $\alpha_e$ is the force-voltage factor in the electromechanical coupling. The shaker delivers a harmonic base excitation with an amplitude $B_f$ and frequency $\omega$. $C_p$ is the clamped capacitance of the piezoelectric patch. The displacement of the equivalent mass is denoted by $x$, and $v_p$ denotes the voltage of the piezoelectric element. $i_p$ represents the current flowing through the interface circuit. In the circuit, $C_b$ bears dual functions of voltage bias-flip and energy storage. An inductor $L$ with an equivalent series resistance (ESR) $r$ facilitates the charge manipulation together with the diode and MOSFET network. System parameters are summarized in Table \ref{tab:para}.

\subsection{System Characterization}
\label{subsec:circuit}

To quantify the electrically induced dynamics by BECC, we first formulate ${Z_e}$ the electrical equivalent impedance of the $C_p$ and BECC combination.  The current flowing through the $C_p$ and BECC combination is indicated as follows:
\begin{equation}
    i_{h}\left(t\right) = \alpha_e \dot{x}(t),
    \label{eq:i_p}
\end{equation}
where $i_h$ is a periodic current with the possibility of higher-order harmonics due to the presence of cubic nonlinearity in \eqref{eq:eom}.  For a linear energy harvester, $i_h$ only contains the fundamental harmonic, and voltage flips occur at current crossing zero points (displacement extremes). While cubic nonlinear energy harvesters span from periodic to chaotic oscillations, these high-order harmonics and chaotic oscillations may induce multiple voltage flips in one fundamental oscillation cycle. To prevent these behaviors, we employ a switch resting time after each voltage flip and assume weak nonlinearity in the system for periodic solutions.  Therefore voltage $v_p$ could still have a first-order resonant period, which can be formulated with a piecewise equation as follows:
\begin{equation}
    {v_p}\left(t \right) = \left\{ 
    \begin{aligned}
        & \frac{1}{C_p} \int_{0}^{t} i_h(t) \di t - {V_M},&
            0 \le  t < \frac{\pi}{\omega};\\
        &{V_M} - \frac{1}{C_p} \int_{\frac{\pi}{\omega}}^{t} i_h(t)\di t,
        &\frac{\pi}{\omega}  \le  t  < \frac{2\pi}{\omega}, 
    \end{aligned}
    \right.
    \label{eq:v_p}
\end{equation}
where $V_M$ is the final voltage after the $M^\text{th}$ bias-flip actions in each synchronized instant \cite{zhao2021bidirectional}.  It can be expressed as a function of the bias voltage ${V_b}$ and the open-circuit voltage $V_{oc}$ according to the equations for energy harvesting modes and vibration exciting modes of BECC \cite{zhao2021bidirectional}. With the definition of the open-circuit voltage as the voltage accumulation on the clamped capacitor $C_p$ during a quarter of a vibration cycle, $V_{oc}$ reads:
\begin{equation}
    V_{oc}=\frac{1}{2C_p} \int_{0}^{\frac{\pi}{\omega}} i_h(t)\di t.
\end{equation}

\begin{figure}[!t]
    \centering
    \includegraphics[width=\columnwidth,page=2]{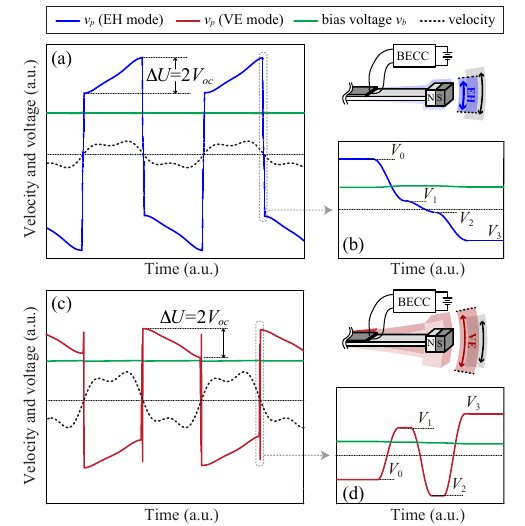}
    \caption{The waveform in different operation modes of BECC. (a) Energy harvesting (EH) mode; (b) The enlarged view of a falling edge showing the bias-flip actions; (c) Vibration exciting (VE) mode; (d) The enlarged view of a rising edge showing the bias-flip actions.}
   \label{fig:wave}
 \end{figure}

In \eqref{eq:v_p}, $V_M<0$ corresponds to energy harvesting modes, and $V_M>0$ corresponds to vibration exciting modes. Different $M$ number indicates the bias-flip action times. In this paper, we choose $M=3$ and realize the S3BF energy harvesting mode and S3BF vibration exciting mode of BECC. These two modes are simply referred to as the energy harvesting (EH) mode and the vibration exciting (VE) mode of BECC. Fig. \ref{fig:wave}(a) and (c) show the voltage and current waveform under the EH and VE modes of BECC. The switching sequence which controls different MOSFETs in Fig. \ref{fig:system} is referenced in \cite{zhao2021bidirectional}. The enlarged view in Fig. \ref{fig:wave}(b) and (d) illustrate the intermediate voltages at a falling edge and a rising edge, which are influenced by the flipping factor $\gamma \in \left(-1,0\right)$ \cite{zhao2021bidirectional}. Under EH mode, the piezoelectric voltage $v_p$ is in-phase with the oscillator velocity $\dot{x}$ (also with piezoelectric current $i_h$), which indicates the energy flow from the mechanical structure into the interface circuit. In contrast, the out-of-phase condition happens under VE mode. It should be noted that $v_p$ is a piecewise and continuous function with respect to the oscillator velocity $\dot{x}$. The piezoelectric coupling force $\alpha_e v_p$ is also a piecewise and continuous force, which does not have a high degree of smoothness. However, here we employ the first-order assumption to avoid the tedious expansion and focus on weak nonlinear monostable systems. A complete solution that addresses higher-order harmonics induced by circuits will be discussed in future work.  

By studying the magnitude and phase relation between the fundamental harmonic of $v_p$ and $i_{h}$, we can formulate the equivalent impedance of the clamped capacitor $C_p$ and BECC combination in the frequency domain \cite{zhao2022positive} as follows:
\begin{equation}
    {Z_{e}}(j\omega ) = \frac{{{V_{p,f}}(j\omega )}}{{{I_h}(j\omega )}} = \frac{1}{{\omega {C_p}}}\left[\frac{4}{\pi }\left(1 - \tilde{V}_M\right) - j\right],
    \label{eq:Ze}
\end{equation}
where $V_{p,f}$ is the magnitude of the fundamental harmonic of $v_p$. $\tilde{V}_M$ is the $V_{oc}$ normalized final voltage after flipping. According to the electromechanical analogy, the electrical impedance can be represented by mechanical parameters as follows \cite{Liang2012}:
\begin{equation}
    \begin{aligned}
        C_{e}&=\alpha_e^{2} \Re{Z_e} \\
          	K_{e}&=-\omega \alpha_e^{2} \Im{Z_e}, 
    \end{aligned}
    \label{eq:CK}
\end{equation}
where $C_e$ and $K_e$ are the electrically induced damping and stiffness. $K_e$ is a constant value since the imaginary part of ${Z_{e}}$ is constant, which represents the added stiffness from $C_p$. On the other hand, the tuning range of $C_e$ depends on the load condition of BECC determined by the intermediate voltages \cite{zhao2020series}. The electrically induced damping range for EH mode is given as follows: 
\begin{equation}
    \left\{C_e \in \mathbb{R}^+ \left| \frac{4\alpha_e^2 (1-\gamma)}{\omega \pi C_p} \le C_e \le \frac{4\alpha_e^2 (3-3\gamma)}{\omega \pi C_p (1+\gamma)} \right.\right\}.
    \label{eq:EH_Ce}
\end{equation}
It can be seen that under the fundamental harmonic assumption, the dependence of $C_e$ on the oscillator velocity is canceled. In EH mode, $C_e$ is positive, which indicates that the nonlinear oscillator is damped for energy harvesting purposes. The stored energy in the bias/storage capacitor can be used for vibration-exciting propose. Therefore, with the maximum bias voltage $v_{b,max}=2V_{oc}(1-\gamma)/(1+\gamma)$ \cite{zhao2020series}, the range of $C_e$ under VE mode reads:
\begin{equation}
    \left\{C_e \in \mathbb{R}^- \left| \frac{-4\alpha_e^2(3\gamma^4-5\gamma^3+8\gamma^2-7\gamma+1)}{\omega \pi C_p(1+\gamma)^2(\gamma^2-\gamma+1)} \le C_e < 0 \right.\right\}.
    \label{eq:VE_Ce}
\end{equation}
In VE mode, $C_e$ is negative. Therefore, adding $C_e$ to mechanical damping $C$ reduces the total damping for higher vibration amplitudes. It could even form a gross effect of negative damping corresponding to an actuation force, which excites the oscillator from a quiescent state \cite{zhao2021bidirectional}. 

\section{Time-sharing Orbit Jumps}
\label{sec:dynamics}
In order to facilitate the nonlinear oscillator to reach the HEO for more harvested energy, the main idea is to perturb the oscillator into the basin of attractions for HEOs, regardless of the method used. Therefore, stability analysis and state-space evaluation are necessary to reveal the mechanism of orbit jumps and the influence of the proposed orbit jump solution by the time-sharing operations of BECC. 

\subsection{Autonomous Case}
\label{subsec:auto}
We first analyze the stability of the nonlinear oscillator without external excitation force, which forms an autonomous system with the equation of motion:
\begin{equation}
	M\ddot{x}+C\dot{x}+Kx-K_1x+K_3x^3+K_ex+C_e\dot{x}=0,
	\label{eq:eom_auto}
\end{equation}
where the electrically induced $C_e$ and $K_e$ can be determined with \eqref{eq:CK} to \eqref{eq:VE_Ce}. By writing \eqref{eq:eom_auto} into state-space form, this nonlinear oscillator can be represented as follows:
\begin{equation}
    \left\{
    \begin{aligned}
        \dot{x}&=y, \\
        \dot{y}&=-\omega_0^2x-k_3x^3-cy,
    \end{aligned}
    \right.
\end{equation}
where $y$ represents the velocity of the oscillator. $\omega_0^2=(K+K_e-K_1)/M$, $k_3=K_3/M$, $c=(C+C_e)/M$ are the mass $M$ normalized gross effect of linear stiffness, nonlinear cubic stiffness, and the gross effect of linear damping, respectively. Since $\omega_0^2$ is a constant parameter above zero, this nonlinear oscillator is a monostable nonlinear oscillator with a fixed point $x_p$ at the origin. The linearization of the dynamics around the fixed point can be determined by the Jacobian matrix:
\begin{equation}
    \boldsymbol J=
    \begin{bmatrix}
    0 & 1\\
    -\omega_0^2-k_3 x^2 & -c
    \end{bmatrix}.
    \label{eq:Jaco1}
\end{equation}

By substituting $x=x_p=0$, we obtain the eigenvalues:
\begin{equation}
    \lambda_{1,2}(C_e)=-\frac{c}{2}\pm \frac{\sqrt{c^2-4\omega_0^2}}{2}.
\end{equation}
Assuming $4\omega_0^2>c^2$, the eigenvalues are a pair of complex conjugates. The sign of their real parts is determined by the sign of $c$. Therefore, the linearized dynamics around the fixed point also depend on the sign of the real part of the eigenvalues. When $c=0$, the two eigenvalues are purely imaginary, indicating that a two-dimensional center manifold will be determined around the origin. However, different from the nonlinear stiffness $K_3$ dependent center manifold, which features a Hopf bifurcation with limit cycles on one side of the origin \cite{guckenheimer2013nonlinear}. There is no limit cycle on either side of the bifurcation point when the damping $C_e$ is chosen as a dependent parameter. 

\begin{figure}[!t]
    \centering
    \includegraphics[width=\columnwidth,page=3]{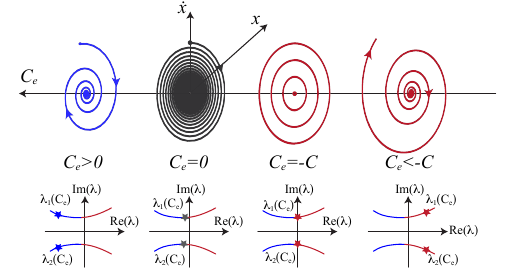}
    \caption{Four different phase portraits correspond to different $C_e$ with the eigenvalues depending on $C_e$.}
   \label{fig:Ce_phase}
 \end{figure}

Fig. \ref{fig:Ce_phase} illustrates four different phase portraits from time-domain integration corresponding to four $C_e$ cases. $C_e=0$ corresponds to the original nonlinear oscillator without BECC, whose phase portrait is a stable spiral with its original damping $C$. Under EH mode, the total damping $C+C_e$ becomes larger; therefore, the phase portrait also shows a stable spiral but decays faster to the origin. When $C_e=-C$ for VE mode, the origin becomes a nonlinear center. This is a special case of a degenerate Hopf bifurcation, where the eigenvalues have purely imaginary parts. As $C_e$ further decreases when $C_e<-C$, the real parts of eigenvalues become positive. This corresponds to an unstable spiral such that a self-excited oscillator is formed to achieve higher vibration amplitudes.

\subsection{Nonautonomous Case}
\label{subsec:noauto}
Besides the autonomous stability analysis, we also explore the frequency response of this nonlinear energy harvester under an external periodic driving force around the primary resonance of the nonlinear oscillator. With the assumptions of weak damping, nonlinear reaction force, and weak external force, the governing equation of the nonlinear oscillator with BECC can be written as follows:
\begin{equation}
    \ddot{x}+ c\dot{x}+\omega_0^2x+k_3 x^3= b_f \cos\left(\omega t \right),
    \label{eq:ms_eom}
\end{equation}
where $b_f=B_f/M$ is the mass normalized force amplitude, and $\omega$ is the base excitation frequency. Note that an analytical approximation for \eqref{eq:ms_eom} takes the following form: 
\begin{equation}
	x(t)=a\left(t\right)\cos\left[\omega t+\varphi \left(t\right)\right]+O\left(\varepsilon\right),
\end{equation}
where $a(t)$ and $\varphi(t)$ are slowly time-varying real-valued amplitude and phase. This solution can be constructed with asymptotic series through the multiple time scales method as \cite{kevorkian2012multiple}:
\begin{equation}
x=x_0(\tau, T)+\varepsilon x_1(\tau, T)+\cdots,
\label{eq:xseries}
\end{equation}
where the fast and slow time variables are defined as $\tau=t$ and $T=\varepsilon t$. Compared with the harmonic balance or direct time-domain integration methods, the multiple time scales method can not only solve transient response given fast dynamics in a nonlinear system, but also increases the numerical stability and efficiency by decomposing the system into different time scales and integrating with different time steps \cite{kevorkian2012multiple}. This method particularly fits the nonlinear energy harvester and orbit jump solution proposed in this paper. The time derivatives of \eqref{eq:xseries} read:
\begin{equation}
\left\{\begin{array}{l}
\frac{\di}{\di t}=\frac{\partial}{\partial \tau}+\varepsilon \frac{\partial}{\partial T}+\cdots =D_0+\varepsilon D_1+\cdots, \\
\frac{\di^2}{\di t^2}=D_0^2+2 \varepsilon D_0 D_1+\varepsilon^2 D_1^2+\cdots.
\end{array} \right.
\end{equation}
A frequency detuning parameter $\sigma$ for the external force is given as follows:
\begin{equation}
    \omega_0^2=\omega^2+\varepsilon \sigma.
\end{equation}
By substituting \eqref{eq:xseries} into \eqref{eq:ms_eom}, one can collect the $\varepsilon^0$ and $\varepsilon^1$ items as follows:
\begin{equation}
\left\{\begin{aligned}
D_0^2 x_0+\omega^2 x_0=&0, \\
D_0^2 x_1+\omega^2 x_1=&-\sigma^* x_0-2 D_0 D_1 x_0-c^*D_0 x_0-k_3^* x_0^3\\
&+b_f^*\cos(\omega \tau) ,
\end{aligned}
\right.
\label{eq:2order}
\end{equation}
where $\sigma^*=\sigma/\varepsilon$, $c^*=c/\varepsilon$, $k_3^*=k_3/\varepsilon$, and $b_f^*=b_f/\varepsilon$ are the $\varepsilon$-scaled parameters for nonlinear analysis.

The general solution for the first component of \eqref{eq:xseries} can be written as follows:
\begin{equation}
x_0=A(T) e^{j \omega \tau}+\bar{A}(T) e^{-j \omega \tau},
\label{eq:x0}
\end{equation}
where $A$ and $\bar{A}$ represent complex conjugates. By substituting \eqref{eq:x0} into the second equation of \eqref{eq:2order}, the result is:
\begin{equation}
    D_0^2 x_1+\omega^2 x_1=STe^{j\omega \tau}-k_3^*A^3e^{3j\omega \tau}+cc,
\end{equation}
where $ST$ and $cc$ represent the secular term and the complex conjugate. On setting the source of the secular terms to zero, it gives:
\begin{equation}
2j\omega D_1 A+\sigma^* A+jc^* A \omega+3 k_3^*A^2\bar{A}-\frac{1}{2} b_f^* =0.
\label{eq:st}
\end{equation}
By defining the derivative of amplitude $\di A/\di t=\varepsilon D_1A$ and introducing the polar form of $A=ae^{j\varphi}/2$, \eqref{eq:st} can be separated into real and imaginary parts for the slow flow of $A$ as follows:
\begin{equation}
    \begin{aligned}
        \frac{\di a}{\di t} & =-\frac{a c}{2}-\frac{b_f }{2\omega} \sin \varphi, \\
        \frac{\di \varphi}{\di t} & =\frac{\omega_0^2-\omega^2}{2 \omega}+\frac{3 k_3 a^2}{8 \omega}-\frac{b_f }{2\omega a} \cos \varphi.
    \end{aligned}
    \label{eq:slowflow}
\end{equation}
The two derivatives describe the slow flows of the amplitude and phase. Under a steady state, the fixed points or the amplitudes and phase can be obtained by equating the right-hand side to zero as follows:
\begin{equation}
    \begin{aligned}
        &\frac{a c}{2}=-\frac{b_f }{2\omega} \sin \varphi, \\
        & \frac{a\left(\omega^2-\omega_0^2\right)}{2 \omega}-\frac{3 k_3 a^3}{8 \omega}=-\frac{b_f }{2\omega} \cos \varphi .
    \end{aligned}
    \label{eq:fixps}
\end{equation}
Squaring and adding the two equations in \eqref{eq:fixps} yields the frequency response function of the nonlinear energy harvester:
\begin{equation}
   \frac{a}{b_f}=\frac{1}{\sqrt{c^2\omega^2+\left(\omega^2-\omega_0^2-\frac{3}{4}k_3 a^2\right)^2}}.
   \label{eq:frf}
\end{equation}

\begin{figure}[!t]
    \centering
    \includegraphics[width=\columnwidth,page=4]{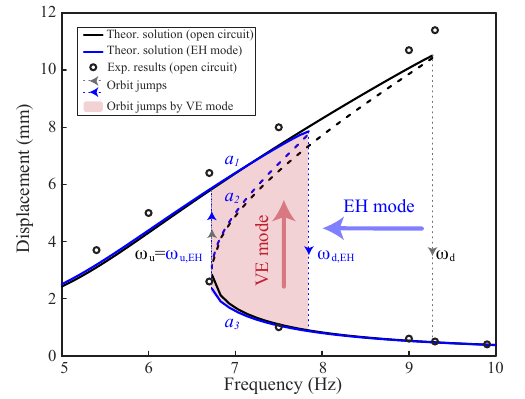}
    \caption{Theoretical and experimental frequency response of the nonlinear energy harvester. The black lines represent the open-circuit case. The blue lines represent the energy harvesting case. The red shadow region indicates the hysteresis region where orbit jumps can be realized from LEO to HEO.}
    \label{fig:frf}
\end{figure}

Alternatively, \eqref{eq:ms_eom} can also be solved by using the harmonic balance method \cite{zhao2020dual}. The expression of the solution has the same form as \eqref{eq:frf} with first-order harmonic assumption. The steady-state displacement amplitude of the nonlinear energy harvester is shown in Fig. \ref{fig:frf}. It can be seen that, due to the cubic nonlinear stiffness, the frequency response has a hardening effect with the presence of a hysteresis range. The two stable fixed points from \eqref{eq:fixps} and an unstable fixed point give rise to the upper and lower branches, which forms the hysteresis range between the two gray dash lines in Fig. \ref{fig:frf}.

In order to quantify the influence of electrically induced damping $C_e$ on the hysteresis range of the nonlinear harvester, the two critical frequency $\omega_{u}$ and $\omega_{d}$ for up and down orbit jumps can be determined by imposing the derivative $\di\omega/\di a=0$ in \eqref{eq:frf}. Under the weak damping assumption \cite{brennan2008jump}, the two critical frequencies read:  
\begin{align}
    \omega_{u}&=\omega_0\left[1+\frac{1}{2}\left(\frac{3}{2}\right)^{4/3}\left(\frac{k_3 b_f^2}{\omega_0^6}\right)^{1/3}\right],
    \label{eq:omega_u}\\
    \omega_{d}&=\frac{\omega_0}{\sqrt{2}}\left[1+\left(1+\frac{3k_3b_f^2}{\omega_0^4 c^2}\right)^{1/2}\right]^{1/2}.
    \label{eq:omega_d}
\end{align}

\begin{figure}[!t]
    \centering
    \includegraphics[width=\columnwidth,page=5]{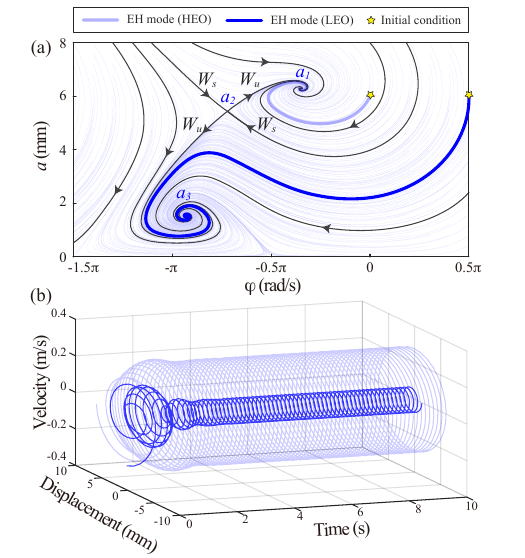}
    \caption{Phase portrait and phase evolution under EH mode with base excitation frequency at 7 Hz. (a) $(a,\varphi)$ phase portrait; (b) Displacement and velocity $(x,\dot{x})$ phase evolution over time. }
    \label{fig:EHphase}
\end{figure}
 
\begin{figure*}[!t]
    \centering
    \includegraphics[width=7in,page=6]{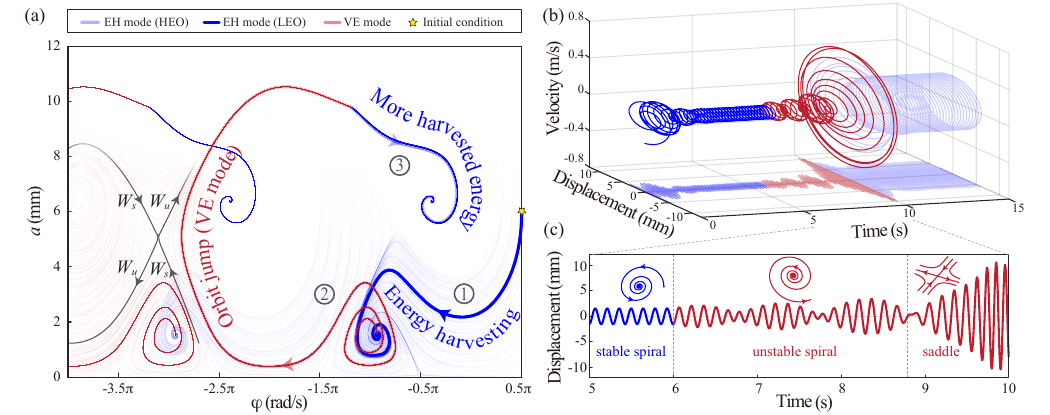}
    \caption{Phase portrait and phase evolution of orbit jumps. The system starts from EH mode to VE mode and then to EH mode again under a 7 Hz base excitation. (a) $(a,\varphi)$ phase portrait with trajectories for orbit jumps; (b) Displacement and velocity $(x,\dot{x})$ phase evolution and its displacement projection over time; (c) The detailed phase evolution of a trail of orbit jump with its stability types.}
    \label{fig:jump}
\end{figure*}
 
It can be seen that $\omega_{u}$ does not depend on the damping coefficient while $\omega_{d}$ does. This dependence on damping leads to the decrease of $\omega_{d}$ from an open-circuit case to an energy harvesting case due to a more significant gross effect of total damping $c$ as shown in Fig. \ref{fig:frf}. Therefore, orbit jumps from low-energy orbits to high-energy orbits take effect in the hysteresis range of a nonlinear harvester under EH mode indicated by the red shaded area in Fig. \ref{fig:frf}.

When a slowly varying parameter $C_e$ is introduced, the frequency-domain analysis based on the steady-state assumption in \eqref{eq:fixps}, like harmonic balance, are no longer applicable. Thus, in this paper, we utilize the state-space equations in \eqref{eq:slowflow} to better illustrate the influence of varying electrically induced damping $C_e$ with a periodic harmonic excitation force. The stability of the fixed points and solution branches of the nonlinear oscillator can be determined by the Jacobian of the slow flows in \eqref{eq:slowflow} as follows:
\begin{equation}
    \boldsymbol J=
    \begin{bmatrix}
    -\frac{c}{2} & a\left(\frac{\omega^2-\omega_0^2}{2\omega}-\frac{3k_3a^2}{8\omega}\right)\\
    -\frac{1}{a}\left(\frac{\omega^2-\omega_0^2}{2\omega}-\frac{9k_3a^2}{8\omega}\right) & -\frac{c}{2}
    \end{bmatrix}.
    \label{eq:Jaco2}
\end{equation}
The eigenvalues of the Jacobian can be solved as follows:
\begin{equation}
    \lambda_{1,2}=-\frac{c}{2}\pm \sqrt{\left(\frac{\sigma_\varepsilon}{2\omega}-\frac{3k_3a^2}{8\omega}\right)\left(\frac{9k_3a^2}{8\omega}-\frac{\sigma_\varepsilon}{2\omega}\right)},
\end{equation}
where $\sigma_\varepsilon=\omega^2-\omega_0^2$. For EH mode, the gross effect of damping $c<0$. The stability of three branches in the hysteresis range indicated with $a_1$, $a_2$, and $a_3$ in Fig. \ref{fig:frf} are determined by their eigenvalues. Taking an example, where $\omega=2\pi7$ (rad/s), the two stable fixed points $a_1$ and $a_3$ both have a pair of complex conjugate eigenvalues with negative real parts, which create stable spirals. Another unstable fixed point $a_2$ has a negative and a positive real eigenvalue, which forms a saddle point. As shown in Fig. \ref{fig:EHphase}(a), this saddle point associates a stable manifold $W_s$ and an unstable manifold $W_u$, which are tangent to the corresponding eigenvectors. The stable manifold $W_s$ partitions the phase portrait into two regions, which are the basins of attraction of the stable fixed points $a_1$ and $a_3$. Different initial conditions attract the trajectories to either $a_1$ (high-energy orbit) or $a_3$ (low-energy orbit).

Taking two different initial conditions indicated with yellow stars in Fig. \ref{fig:EHphase}(a), on both sides of the stable manifold $W_s$, they finally rest in different energy orbits. As shown with the time evolution of their $(x, \dot{x})$ phase portraits in Fig. \ref{fig:EHphase}(b), the light blue trajectory for HEO corresponds to the initial condition lies inside the region surrounded by the stable manifold. Therefore, orbit jumps from LEO to HEO equals to shift and tune the $(a, \varphi)$ phase portrait of the oscillators such that by VE mode of BECC, they can finally be attracted to the high-energy orbit or the $a_1$ fixed point.

For VE mode, the two stable fixed points $a_1$ and $a_3$ in EH mode become unstable if $c<0$ is satisfied. If we also assume a small absolute value of negative $C_e$, $a_2$ will still remain a saddle point. The two unstable spirals and the saddle node are enclosed by red trajectories in Fig. \ref{fig:jump}(a). We first take the initial conditions for LEOs and carry on the orbit jump process. Fig. \ref{fig:jump}(a) shows their phase evolution under different circuit modes. The three steps for orbit jumps labeled with numbers are elaborated as follows:
\begin{enumerate}[itemsep=2pt,label=\protect\mycirc{\arabic*}]
    \item The oscillator is first attracted to the fixed point for LEO, and the circuit is initially in EH mode to harvest energy with a positive $C_e$. The trajectories for this step are stable spirals. All the trajectories attracted to LEO have the same $a$ and $\varphi$ at steady states. The $(x, \dot{x})$ time evolution for energy harvesting at the LEO is shown with blue curves in Fig. \ref{fig:jump}(b).
    \item The circuit is switched to VE mode with negative $C_e$ for vibration exciting.  The oscillator first leaves the unstable fixed point in the $(a, \varphi)$ phase portrait in a reverse spiral manner. When trajectories meet the saddle point, they will be first attracted and then repelled by the saddle point into higher displacement amplitudes. The $(x, \dot{x})$ time evolution for vibration exciting is shown with red curves in Fig. \ref{fig:jump}(b).
    \item The circuit is switched back to EH mode. The trajectories right now lie in the basin of attraction for the HEO. Therefore, they finally rest in the HEO for more harvested energy. The $(x, \dot{x})$ time evolution in energy harvesting at the HEO is shown with light blue curves in Fig. \ref{fig:jump}(b).
\end{enumerate}

\begin{figure}[!t]
    \centering
    \includegraphics[width=\columnwidth,page=7]{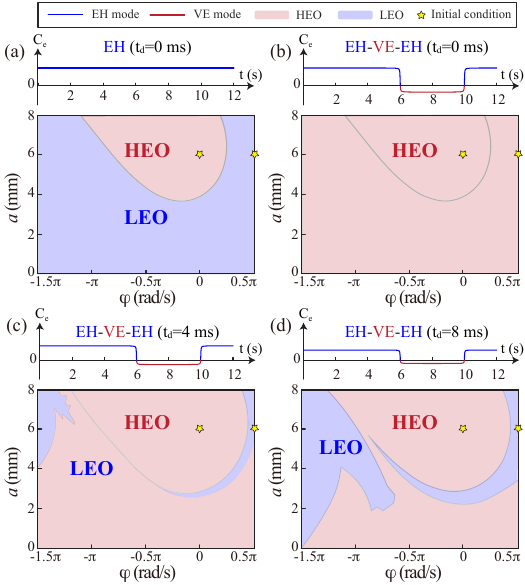}
    \caption{The basin of attractions for EH mode and orbit jumps (EH-VE-EH) under different initial conditions and time delays. (a) EH mode with no time delay; (b) Orbit jumps (EH-VE-EH) with no time delay; (c) Orbit jumps (EH-VE-EH) with time delay $t_d=4$ ms; (d) Orbit jumps (EH-VE-EH) with time delay $t_d=8$ ms.}
   \label{fig:ini}
 \end{figure}

By closely observing the $(x, \dot{x})$ time evolution of orbit jump in Fig. \ref{fig:jump}(c), the displacement envelop depicted by the red curve under the VE mode becomes almost periodic, which is similar to a beat. This is actually due to the phase shift in the $(a, \varphi)$ phase portrait. At the beginning of vibration exciting, the trajectory slowly leaves the unstable fixed point in an unstable spiral manner, during which the displacement amplitude begins to grow. After several cycles, the saddle point attracts and repels the trajectory with the stable and unstable manifolds. During this period, there exist a few cycles of increase and decrease of displacement amplitude, which forms the beating waveform. However, the envelope of the displacement always has an increasing trend due to the negative damping effect. After repelled by the saddle node, the displacement amplitude of the oscillator is amplified and lies in the basin of attraction for the HEO. Therefore, switching the circuit back to EH mode allows the trajectory to be easily attracted to another stable fixed point.

Fig. \ref{fig:ini} shows the evolution of $C_e$ with respect to time and the basin of attractions with different initial conditions.  For the EH mode of BECC, as shown in Fig. \ref{fig:ini}(a), $C_e$ maintains a constant positive value which corresponds to a certain load condition for energy harvesting. For the transition from EH to VE mode, $C_e$ turns from a positive to a negative value starting from $6$s to $10$s. A smooth transition is applied to avoid undesired numerical instability in the simulations. It can be seen from  Fig. \ref{fig:ini}(a) that, for EH mode, the basin of attractions for HEO is surrounded by the stable manifold $W_s$. While with the phase manipulation by VE mode, the basin of attractions for HEO is extended to all initial conditions in Fig. \ref{fig:ini}(b). The above discussions of the basin of attractions are based on the ideal model that the system does not have time delays. However, in real applications, the laser vibrometer and microcontroller, which form a feedback control loop, may introduce small time delays. The time delay may affect the behavior of a nonlinear energy harvester and its orbit jump. Thus, it is necessary to investigate the effect of time delay in the energy harvesting system. A delay time $t_d$ is therefore introduced to the piezoelectric voltage $v_p$ in \eqref{eq:v_p} as follows:
\begin{equation}
    {v_p^d}\left(t \right) = \left\{ 
    \begin{aligned}
        & \frac{1}{C_p} \int_{\frac{\beta}{\omega}}^{t} i_h\di t - {V_M},&
            \beta \le  \omega t < \pi+\beta;\\
        &{V_M} - \frac{1}{C_p} \int_{\frac{\pi+\beta}{\omega}}^{t} i_h\di t,
        &\pi+\beta  \le  \omega t  < 2\pi+\beta, 
    \end{aligned}
    \right.
    \label{eq:v_p_b}
\end{equation}
where $\beta=\omega t_d $ represents a delay phase satisfying $0<\beta<\pi/2$. Following the same procedures mentioned in Subsection \ref{subsec:circuit}. The equivalent impedance of BECC with a delay time reads:
 \begin{equation}
 \begin{aligned}
         {Z_{e}^d}\left(j\omega,\beta\right) &= \frac{4}{\pi{\omega {C_p}}}\left[\cos\beta\left(\cos\beta-\tilde{V}_M\right)\right.\\
         &+\left.j\left(\sin\beta-\sin\beta \cos\beta-\frac{\pi}{4}\right)\right].     
 \end{aligned}
    \label{eq:Zed}
 \end{equation}
 
With the presence of the time delay, the absolute value of the real part of $Z_{e}^d$ becomes smaller compared with that in \eqref{eq:Ze}. Therefore, the absolute value of the electrically induced damping $C_e$ also becomes smaller. It not only reduces the gross effect of damping $c$ and harvested power but also decreases the effect of vibration exciting in VE mode.  Take a bias voltage $v_b=2V_{oc}$ as an example. The basin of attractions corresponding to different delay times are illustrated in Fig. \ref{fig:ini}(b) to (d). It can be seen that some initial conditions achieved HEO from LEO with no time delay turn back to LEO eventually in Fig. \ref{fig:ini}(c). This phenomenon is more prominent in Fig. \ref{fig:ini}(d) when $t_d=8$ ms. Therefore, a larger bias voltage is needed to increase the electrically induced damping and achieve full-hysteresis-range orbit jumps under the same delay time.

    \begin{table}[t!]
    \caption{System parameters}
     \footnotesize
     \label{tab:para}
    \begin{center}
    \setlength{\tabcolsep}{1mm}{
    \begin{tabular}{llll}
    \toprule
    \multicolumn{4}{l}{\textbf{Harvester geometry:}}   \\
    \midrule
    Cantilever (mm$^3$) & 90$\times$10$\times$1  & Material  & Copper \\
    Magnet (mm$^3$) &  10$\times$10$\times$10 & Material  & Neodymium \\
    Piezoelectric patch (mm$^3$) &  56$\times$7$\times$0.2 & Material  & PZT \\
    \multicolumn{4}{l}{Center-to-center distance of dipole magnets (mm): 32.3}   \\
        \midrule
    \multicolumn{4}{l}{\textbf{Mechanical parameters:}}   \\
        \midrule
    $M$ (g)	& 8.9 & $C$ (Ns/m) & 0.011 \\
    $K$ (N/m) & 52.6 & $K_1$ (N/m)	& 41.46 \\
    $K_2$ (kN/m$^3$) &	226.7 & $B_f$ (mN) & 8.9 \\
        \midrule
    \multicolumn{4}{l}{\textbf{Electrical parameters:}}  \\
        \midrule
    $\alpha_e$ (mN/V) & 0.127 & $\gamma$ & $-$0.38  \\
    $C_p$ (nF)	& 35.31 & $L$ (mH) & 47  \\
    $r$ ($\Omega$) & 45.2 & $C_b$ ($\mu$F) & 47 \\ 
    $R_p$ (k$\Omega$) & 870 & MOSFET & ZVN(P)4424 \\
    \bottomrule
    \end{tabular}}
    \end{center}
    \end{table}


\section{Experiment}
\label{sec:exp}

\begin{figure*}[!t]
    \centering
    \includegraphics[width=2\columnwidth,page=8]{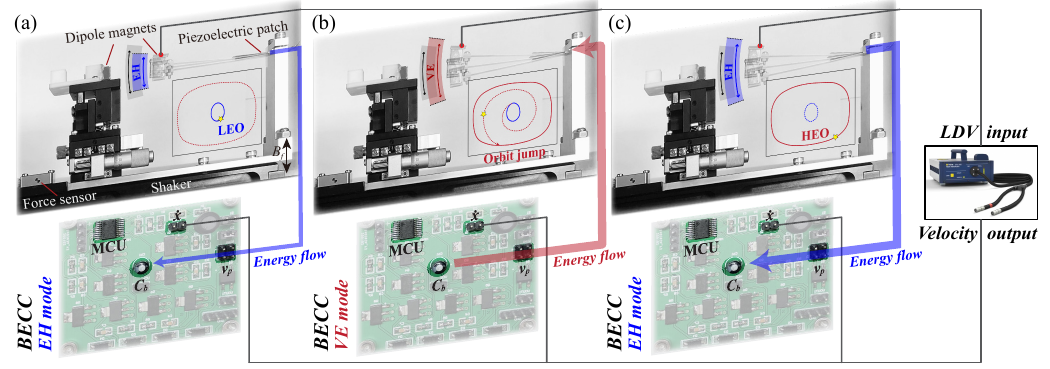}
    \caption{Experimental setup and the orbit jump setup. The blue and red shadow lines indicate the energy flows for energy harvesting and vibration exciting. (a) Energy harvesting at LEO with low output power; (b) Orbit jump with vibration exciting mode; (c) Energy harvesting at HEO with high output power.}
    \label{fig:setup}
\end{figure*}
 
In Fig. \ref{fig:frf}, the frequency response function of the nonlinear harvester has a hysteresis region between 6.7 Hz and 7.8 Hz. This wide span makes the nonlinear system surpass the linear system in terms of bandwidth. However, there is a premise for making the broadband effect. Because of the dual roots within this hysteresis region, one in LEO and the other in HEO, orbit jump capability is necessary to ensure a high power output. The BECC provides a flexible control to manipulate the energy flow between the mechanical and electrical ends. With the help of the vibration exciting mode provided by BECC, it is possible to realize the orbit jump using the same switched-mode energy harvesting circuit. This section validates this self-contained time-sharing orbit jump and energy harvesting solution. 

\subsection{Setup}

The experimental setup is illustrated in Fig. \ref{fig:setup}. A personal computer (PC) controls the base excitation of the energy harvester through a shaker system (APS420, SPEKTRA). A constant base excitation inertial force $B_f=8.9$ mN is applied to the clamped end of the cantilevered harvester under a constant acceleration magnitude of $1$ m/s$^2$. A laser Doppler vibrometer (OFV-552/5000, Polytec) monitors the tip displacement and velocity of the oscillator. It sends the information to the PC and a microcontroller (MSP430G2553, Texas Instrument) as an interrupt signal for synchronizing the BECC switch controls. The piezoelectric patch is connected with BECC following a bias/storage capacitor $C_b$ of 47 $\mu$F for both energy harvesting and vibration exciting functions. An oscilloscope (HDO6104A, Teledyne) tracks the operation waveform of this nonlinear energy harvesting system. The detailed geometric, mechanical, and electrical parameters of this monostable energy harvester are listed in Table \ref{tab:para}. The orbit jump setup is also explained in Fig. \ref{fig:setup}. As the system is entrained on LEO, as shown in Fig. \ref{fig:setup}(a), the energy harvester collects energy from the vibration source and stores the energy in the bias/storage capacitor at a low rate. As the bias voltage in capacitor $C_b$ reaches 30 V, 
The BECC is switched to vibration exciting mode. The stored energy is now boosted back into the mechanical structure to amplify the mechanical vibration, as illustrated in Fig. \ref{fig:setup}(b). When the oscillator gradually reaches HEO, BECC is switched back to energy harvesting mode. As a result, the system stabilizes on HEO and realizes higher power output, as shown in Fig. \ref{fig:setup}(c).

\subsection{Results}

\begin{figure}[!t]
    \centering
    \includegraphics[width=\columnwidth,page=9]{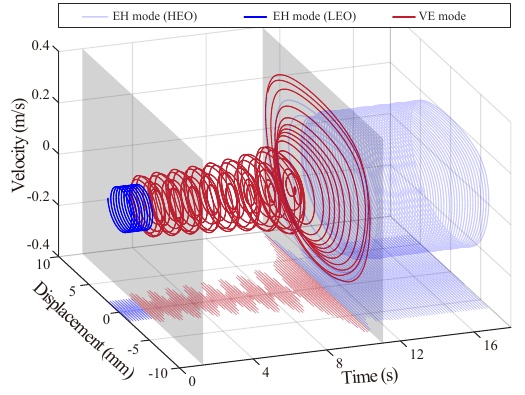}
    \caption{A trial of experimental orbit jump in a monostable nonlinear energy harvester with BECC. The curves represent the displacement and velocity $(x,\dot{x})$ phase evolution over time. The orbit-jump segment under VE mode is illustrated in red within two vertical gray planes, which indicate the mode transitions.}
   \label{fig:mono_p}
 \end{figure}

With the setup and proposed working mechanism, orbit jumps at different base excitation frequencies are carried out experimentally. One of the experimental trials is shown in Fig. \ref{fig:mono_p}. The magnitude and frequency of the sinusoidal base excitation are 1 m/s$^2$ and 7 Hz, respectively. The frequency falls in the hysteresis region of the nonlinear oscillator. The red curves note that BECC is under the vibration exciting mode, while the blue curves represent the energy harvesting mode. Two vertical gray planes indicate the mode-transition instants. As we can see from the figure, in the first segment, the system stays at LEO to collect energy at a low rate. At 1.6 s, the vibration exciting mode of BECC is turned on by the microcontroller on the PCB board of BECC. The vibration-exciting actions send energy back from the bias/storage capacitor $C_b$ to the oscillator to provoke large oscillations. Accordingly, the oscillator vibration amplitude grows gradually and jumps over HEO by the saddle node. Afterward, the BECC is switched back to the energy harvesting mode. The system remains on HEO and harnesses energy at a larger power. During the vibration exciting process, the system experiences the oscillation patterns mentioned above in Fig. \ref{fig:jump}, as the projected displacement trajectory shows. The gradual increase of displacement amplitude due to the unstable spiral and the saddle node together accounts for the success of orbit jumps. A video clip (\textit{Orbitjump.mp4}) recording the orbit jumps of the energy harvester is attached with the paper. 

\begin{figure}[!t]
    \centering
    \includegraphics[width=\columnwidth,page=10]{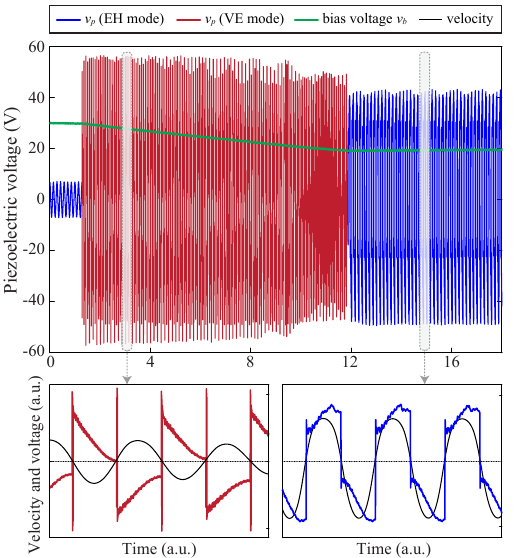}
    \caption{Piezoelectric voltage in the experiment, which corresponds to the orbit jump process described in Fig. \ref{fig:mono_p}. The enlarged views show the experimental piezoelectric voltage in VE mode (red) and EH mode (blue), respectively.}
   \label{fig:mono_v}
 \end{figure}

In one of our previous conference papers \cite{wang2019orbit}, it has been shown that the VE mode of BECC can realize orbit jumps by utilizing the energy stored in the bias/storage capacitor. Here we take a detailed evaluation of energy consumption and recovery. As mentioned above, the piezoelectric voltage is captured and shown in Fig. \ref{fig:mono_v} for an orbit jump trial. The peak voltage for the EH mode has been amplified several times after the orbit jump. During vibration exciting, the exciting voltage gradually decreases with voltage drop on the bias/storage capacitor. The enlarged views show the nonlinear oscillator's detailed voltage and velocity waveform for VE and EH modes, respectively.  The out-of-phase and the in-phase relationships between piezoelectric voltage and velocity correspond to the energy flow from the electrical to the mechanical domain and reverse directions, respectively, in the nonlinear harvesting system. The energy flow directions agree with the voltage drop and rise across $C_b$ under different operation modes.

The energy gain and loss of the bias/storage capacitor $C_b$ are studied to evaluate the energy consumption of orbit jumps.  In the whole orbit-jumping process, the energy is consumed in the microcontroller for switching control, vibration excitation, and also in dissipation in parasitic resistance. Neglecting the minor expenditure in the microcontroller \cite{zhao2020series}, the energy consumption for the orbit jump can be calculated as follows:
\begin{equation}
    E_{VE}=\frac{1}{2}C_b\left(V_{bf}^2-V_{af}^2\right)\approx11.2 \text{ mJ},
    \label{eqn.energyVE}
\end{equation}
where $V_{bf}$, $V_{af}$ represent the voltage before and after the exciting action on $C_b$. $V_{bf}$ is pre-charged to 29 V in this case, and $V_{af}$ is measured to be 19.1 V. Given that $C_b=47$ $\mu$F, the circuit consumes about 1.09 mW average power for the vibration exciting propose during the orbit jump process. It should be noted that the energy consumed from $C_b$ will be partially dissipated by the equivalent series resistance of the inductive branch in the bias-flip actions. Thus, the net injected energy from electrical to mechanical domain is smaller than that is extracted from the storage capacitor $C_b$. The detailed electrical performance, energy flow analysis, and efficiency under different operation modes were documented in \cite{zhao2021bidirectional,zhao2022positive}. After orbit jumps, the energy consumed will be charged back with a higher harvested power on HEO. In the absence of a load, the average charging rate of the bias/storage capacitor is regarded as the power output. Before the orbit jumps, the harvested power is 0.011 mW. On HEO, after orbit jumps, the harvested power has been boosted for 9.1 times to 0.1 mW. Under this rate, the consumed energy by the vibration exciting action will be recovered in about 2 minutes. Considering the size of the transducer, there is room further to optimize the system power output and recovery time. The independence of external devices and the compact dual functions of high-capability energy harvesting and vibration excitation demonstrate the advantages of the proposed self-contained solution for time-sharing energy harvesting and orbit jumps. 
Besides the base excitation at 7 Hz, multiple experimental trials are performed under different base excitation frequencies in the hysteresis region. Successful orbit jumps are observed between 6.7 Hz to 7.8 Hz, covering the entire hysteresis band in EH-mode operation. 
Without utilizing extra mechanical or electrical energy sources for vibration excitation, compared with existing studies in literature \cite{Erturk2011, Kitamura2018}, the proposed self-contained solution realizes the orbit jumps within the full hysteresis range under EH-mode.

\section{Discussions}
\label{sec:dis}

 \begin{figure}[!t]
    \centering
    \includegraphics[width=\columnwidth,page=11]{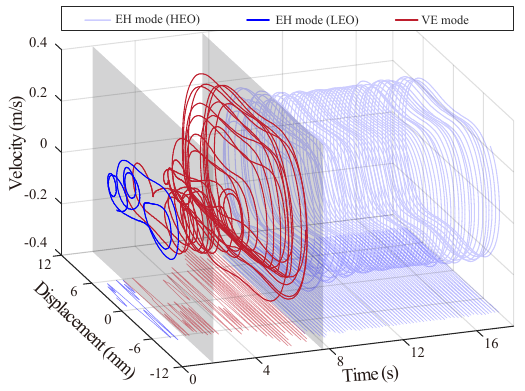}
    \caption{A trial of experimental orbit jump in a bistable nonlinear energy harvester with BECC. The curves represent the phase evolution of the displacement and velocity pair $(x,\dot{x})$ over time. The orbit-jump segment under VE mode is illustrated in red within two vertical gray planes, which indicate the mode transitions.}
   \label{fig:bi_p}
 \end{figure}

In this paper, a monostable nonlinear system is conducted by tuning the distance between the dipole magnets, which expands the bandwidth and increases the output power of the harvester. Besides, other nonlinear systems such as bistable, tristable, and multistable types are extensively explored for similar advantages \cite{fang2022multistability}. However, it is a general case where these nonlinear systems may be trapped in one of its multiple potential wells or randomly transited among wells, which reduces the controllability of orbit jumps. Attempts are carried out with a bistable energy harvester to investigate the feasibility of the proposed orbit jump solution regarding preceding systems. By decreasing the distance between the two repelling magnets such that making $K-K_1<0$, the original single potential well bifurcates into two. Under certain excitation and initial conditions, the bistable harvester can present complex motions \cite{guckenheimer2013nonlinear}.
\begin{itemize}
    \item The interwell oscillation. In this case, the vibrator overcomes the energy barrier of potential wells and travels between two wells. It is referred to as the HEO vibration.
    \item The intrawell oscillation. In this case, the system is trapped in one of its potential wells.
    \item The chaotic motion. The oscillator randomly vibrates without a deterministic path.
\end{itemize}
In case the system does not vibrate on HEO, the BECC activates the VE mode to carry out an orbit jump.

In the experiments, the chaotic motion is mostly observed. The system firstly vibrates chaotically under a sinusoidal base excitation of 5.4 Hz, as shown in Fig. \ref{fig:bi_p}. Then, the oscillator is excited under VE mode for 5.9 s, whose phase portrait is illustrated in red. Afterward, the harvester jumps to HEO, forming the interwell oscillation. Unlike the monostable case, the motion of the bistable oscillator under VE mode does not share the same phase evolution as that in the monostable case, whose phase portrait is shown in Fig. \ref{fig:mono_p}. The first reason is that the intrinsic chaotic motion of a bistable oscillator makes it hard to predict and control the trajectory of a nonlinear oscillator; The second is that the aperiodic motion violates the periodic assumption for calculating the electrical equivalent impedance. Thus the quantitative methods are inappropriate for orbit jumps of bistable energy harvesters with chaotic motions. However, for those bistable energy harvesters whose periodic motions still dominate the dynamics, the methods proposed in this work are still valid.
As a result, the VE mode can only realize orbit jumps around $\omega_u$, where the energy barrier is relatively low. The success of orbit jumps is thus considered the consequence of an abrupt voltage stimulus by the vibration exciting of BECC under the low energy barrier cases between LEO and HEO. It is believed that an abrupt high voltage excitation is preferable for orbit jumps of a chaotic system \cite{Mallick2016}. For other tristable and multistable energy harvesters whose potential barriers are believed to be minor \cite{fang2022multistability}, high-voltage stimuli would be a constructive option. The proposed dual functions of BECC might inspire more efficient energy injection concepts for orbit-jump solutions with nonlinear energy harvesters. 

\section{Conclusion}
\label{sec:conclusion}

In summary, this work integrated a nonlinear piezoelectric energy harvesting with the bidirectional energy conversion circuit (BECC) to realize a time-sharing orbit jump and energy harvesting solution. Based on BECC, we first demonstrated the dual functions of energy harvesting and vibration exciting using a BECC without extra energy sources and actuators.  Then, the ranges of electrically induced parameters under different operation modes were studied with impedance analysis. Furthermore, the stability analysis, frequency response, and state-space phase evolution of the autonomous and nonautonomous systems were performed to analyze the influence of BECC over the dynamics of a nonlinear oscillator. Particularly, the detailed steps for the time-sharing orbit jump from low-energy orbit to high-energy orbit using BECC were studied.  We highlighted the effects of the unstable spiral and saddle node within the nonlinear system. From a practical perspective, the effect of switch time delay on the basin of attractions has also been discussed. Finally, experiments were carried out to validate the feasibility and capability of the proposed orbit jump solution over the entire hysteresis range of the nonlinear energy harvester. The energy evaluation showed that the system’s output power yields a nine-fold increase in the high-energy orbit.  The application in a bistable energy harvesting system also showed the versatility of this solution.   This time-sharing orbit jump solution could facilitate the practical applications of nonlinear energy harvesters. It provides an effective method to attain high-capability piezoelectric energy harvesting on high-energy orbits without the need for extra energy sources and actuators.



\printcredits

\section*{Acknowledgment}

This work was supported by the National Natural Science Foundation of China (Project No. 62271319 and U21B2002) and the Natural Science Foundation of Shanghai (Project No. 21ZR1442300). 
The support for Mr. Bao Zhao provided by the H2020 FET-proactive project METAVEH under the grant agreement 952039, and the ETH Research, Switzerland Grant (ETH-02 20-1) is also acknowledged.

\bibliographystyle{model1-num-names}

\bibliography{cas-refs}




\end{document}